# An Agent-Based Simulation Model for Optimization of the Signalized Intersection Connected to Freeway On-Ramp

Xuejin Wen[1]

[1] Ohio State University; E-Mail: wen036@gmail.com

**ABSTRACT**

Unlike most existing studies on off-ramp traffic signal control, this paper focuses on the optimization problem of the signalized intersection connected to freeway on-ramps. Conflicts are often observed between the traffic heading to an on-ramp and the traffic continuing straight which leads to issues such as intersection overflow, increased delay, and concerns about safety. For studying this problem, a real-world signalized intersection in Buffalo, New York was chosen, which has two through lanes and one short shared (through and right-turn) lane. At the downstream of the intersection are two following on-ramps, one to the highway I-290 West and the other to I-290 East. During peak hours, the shared lane often observes a long queue, which furthermore blocks the through traffic on the parallel lane. To solve this problem, a VISSIM agent-based simulation model was built and calibrated based on field observations. Three potential optimization solutions were proposed and tested with the help of VISSIM: (1) increasing the length of the short shared through and right-turn lane; (2) making the short shared through and right-turn lane right-turn only, and (3) adding a new diverge lane for the right-turn vehicles. According to the simulation results, solution (3) performs the best, resulting in the least vehicle delay time.

# INTRODUCTION

Intersections controlled by traffic signal lights play a major role in road transportation system. Since the first manually operated traffic signal in the 19th century (Mueller, 1970), the topic has been extensively studied. One important principle of signal light optimization is to maximize the throughput and decrease the traffic accident risks simultaneously.

Intersections are one of the core elements of a transportation network. How to optimize the signal timing of one isolated intersection has been extensively studied. For example, Trabia et al. (1999) presented a two-stage fuzzy logic controller for an isolated intersection. In the first stage, it estimates the relative traffic intensities and the second stage determines the extension or termination of the current signal phase (Trabia et al., 1999). Niu et al. (2009) also applied fuzzy control model for the traffic lights at a single intersection.

However, only optimizing one isolated intersection may not solve the optimization problem of the whole transport network. Therefore, the coordinated control system with relevant intersections has attracted more public attention. Split, cycle, and offset optimization technique (SCOOT) is a coordination system in England which can analyze the data from vehicle detectors, predict, and adjust the signal timings to minimize congestion (Hunt et al., 1981). The integrated control of highway off-ramp and its downstream intersection is also necessary to mitigate the congestion. For example, Günther et al. (2012) proposed a methodology which can determine the flow of competing vehicles in the freeway to be detoured to underutilized local roads to improve the systems' capacity. This is extremely useful for the traffic management during the inclement weather (Lin et al., 2015) and traffic accidents (Lin et al., 2016). Li et al. (2016) built mathematical models to study the effect of left-turn vehicles at a signalized intersection with the presence of pedestrian crossings. Accordingly, the macroscopic parameters of traffic flow, such as average flow rate, speed, and density can be determined.

Besides the signal control optimization problem, the specific intersection optimization is also vital to increase the capacity. Although some reference handbooks like the America Highway Capacity Manual (HCM) have provided default capacity calculation methods for intersections, researchers have shown that the realistic cases are different. Tian and Wu (2006) built a probabilistic model for a signalized intersection with a short right-turn lane and validated the model by traffic simulation because HCM would overestimate the capacity and thus underestimate delay. Wu (2007) built a general capacity model for intersections with short shared lanes based on database generated from simulation package VISSIM. Ring and Sadek (2012) studied signalized intersections with auxiliary through lanes and predicted the lane utilization based on the data collected in Buffalo, New York.

As can be seen, traffic simulation is one of the important ways in intersection studies, through which the alternative traffic signal plans and intersection optimizations can be tested. VISSIM is one popular microscopic simulation tool which has been applied for intersection simulation (Ring and Sadek, 2012; Wu, 2007) and border crossing studies (Lin et al., 2013a; Lin et al., 2013b; Lin et al., 2014a; Lin et al., 2014b) and so on. It can customize intersections according to the user's needs by using a link-connector-based network procedure. It also has a strong ability to simulate the traffic signal controllers (Ring and Sadek 2012; Wu, 2007).

This paper will study an isolated intersection with a short shared through and right-turn lane. More interesting is that the intersection locates at the upstream of two consecutive on-ramps, one to the interstate highway I-290 West and the other to the I-290 East. During peak hours, a long queue observed at the short shared lane blocks the through traffic. After the field data collection, a VISSIM model is built and calibrated. Then a few optimization solutions are tested.

The rest arrangement of the paper is as following. The data collection part will introduce the data collection and give a preliminary analysis. The next model development section talks about how to build the intersection model and sets the traffic signal controller in VISSIM. In the results section, different optimization solutions are tested, and the results are shown. The conclusion and future study are discussed in the last section.

**DATA COLLECTION**

As shown in Figure 1, the intersection of Twin Cities Memorial Highway and W Colvin Blvd locates at the upstream of two on-ramps to I-290 West and I-290 East. There are two through lanes and one short shared through and right-turn lane from the northwest to the southeast direction (referred to hereafter as the left lane, the middle lane, and the short lane separately). The left lane and middle lane connects different streets downstream. One is to Eggert Road, and the other is to the Colvin Road. This short lane is where the long queue forms and spills back to block the through traffic.

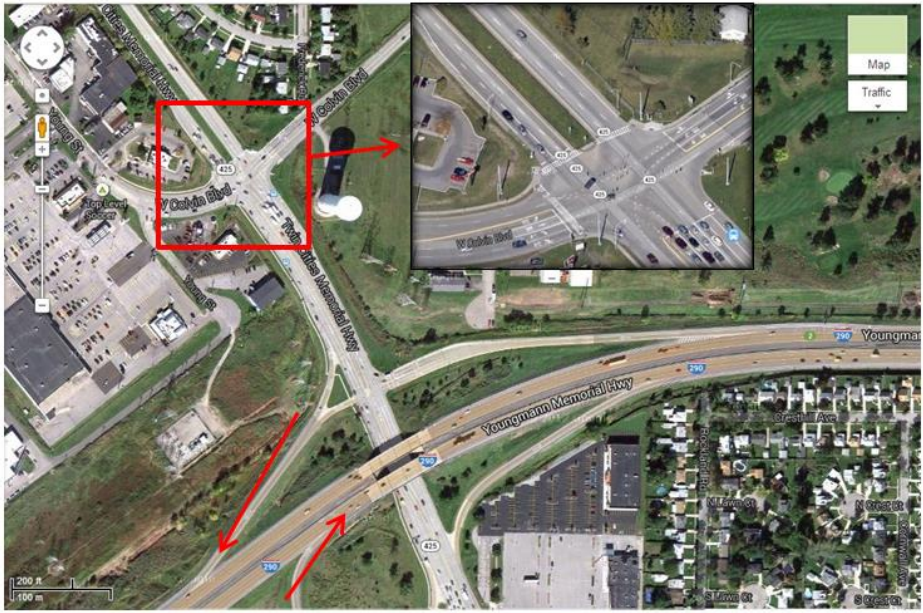

**Figure 1. The intersection in Google map.**

Before building the simulation model, some field observations must be conducted. Firstly, this intersection takes actuated signal control and the minimum green time, maximum green time, yellow time and so on have to be obtained. Secondly, the traffic flow rate for each direction and each lane during peak hour should also be collected. After field recording from 08:16 AM to 09:15 AM of three days 10/16/2013-10/18/2013, Table 1 and Table 2 show the results.

According to the filed observations in Table 1, the traffic signal at this intersection consists of five phases. For Phase 1 the left-turn and through traffic enter the intersection in the same phase from northeast entrance. Phase 2 is similar that the left-turn and through traffic are from southwest entrance. For southeast entrance, the left-turn and through traffic are controlled by two separate signal phases Phase 3 and Phase 4. Finally, the through traffic from northwest entrance is guided by Phase 5. No left-turn traffic is allowed from the northwest entrance.

**Table 1. Traffic Signal Parameters (unit: second)**

| Phase Index | 1 | 2 | 3 | 4 | 5 |
|---|---|---|---|---|---|
| Phase sequence | ⇙ | ⇗ | ↖ | ↗ | ↘ |
| Minimum green | 5 | 5 | 10 | 20 | 20 |
| Maximum green | 10 | 10 | 15 | 30 | 30 |
| Yellow time | 3 | 3 | 3 | 3 | 3 |
| Red clearance | 1 | 1 | 1 | 1 | 1 |

Table 2 shows the lane distribution of traffic flow rate for northwest entrance. As can be seen, more than half of the total traffic flow was going through the short lane. Because of the limitation of experiment equipment, here we only recorded the traffic flow rates for the northwest entrance. For the other entrances, different flow rates will be set as the vehicle inputs of the simulation model, through which it will also be interesting to test the optimization solutions for various scenarios. The details are introduced in next section.

**Table 2. Traffic Flow Rate for Each Lane at Northwest Entrance during Peak Hours 08:16 AM -09:15 AM (unit: pcu/h)**

| Lanes | Northwest Entrance |
|---|---|
| Left Lane | 401 |
| Mid Lane | 343 |
| Short Lane | 836 |
| Total | 1,580 |

## MODEL DEVELOPMENT

### Signal Parameter Setting

With the help of Google Map, the layout of this intersection was built in VISSIM as Figure 2.

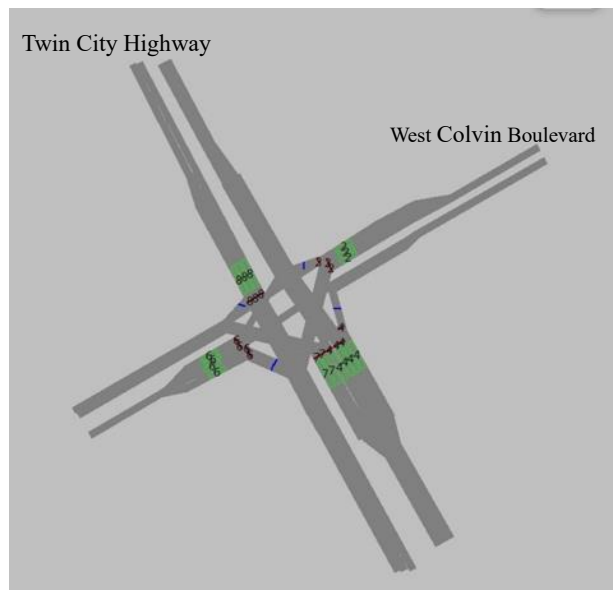

**Figure 2. The intersection in VISSIM.**

For the signal setting, VISSIM takes the Ring Barrier Controller. This controller can simulate fully actuated signal control by defining different signal groups or phases.

Through the field observations, we already know there are five signal phases, in the simulation model they are shown as the red bars in Figure 2. The green rectangles are the detectors for the corresponding signal groups.

The signal parameters including the maximum/minimum green time, yellow time, red clearance time are set according to Table 1. Besides that, the "vehicle extension" in Ring Barrier Controller assigns 2 seconds for West Colvin Boulevard direction and 3 seconds for Twin City Highway direction. The "vehicle extension" extends the green interval time based on the detector status once the phase is green. All five signal groups are flagged with "minimum recall" which means they will receive an automatic vehicle call when they are not green (Mcelroy 2008). At last, the blue bars are stop signs which are used to simulate the right-turn behavior during the red lights.

**Vehicle Input Setting**

For the northwest entrance, the vehicle inputs are the same as the field observations shown in Table 2. Based on the real observations, the right-turn proportion for the vehicles at the short lane is assumed as 10%. For each of the other three entrances, although accurate observations are missing, three traffic flow rates are assumed: low, medium, and high. Meanwhile, the left turn, through, and right turn traffic probabilities are assumed to be equal. The inputs for the simulation model are shown in Table 3.

**Table 3. Assumed Traffic Flow Rates for the Other Three Entrances (unit: pcu/h)**

| Traffic flow rates | Southeast | Northeast | Southwest |
| --- | --- | --- | --- |
| High | 1,500 | 1,000 | 1,000 |
| Medium | 1,000 | 600 | 600 |
| low | 500 | 200 | 200 |

There are 27 possible situations based on the combination of the entrance vehicle inputs. For each situation, the models with different optimizing settings are simulated three times of 3,600 seconds with various random seeds. The results section will show the performances of the optimizing settings.

**RESULTS**

In this paper, three possible optimization solutions are proposed as following: (1) increasing the length of the short lane; (2) making the short shared through and right-turn lane right-turn only; (3) adding a new diverging lane for the right-turn vehicles. For the 27 possible situations, to make it simple, a three-digit number is applied to represent each situation; the three digits represent the traffic flow rates of southeast, northeast, and southwest entrances in order. Each digit has three values, ("1"-high, "2"-"medium", and "3"-"low"). So for example, "112" means the traffic flow rates of southeast and northeast entrances are high, and the flow rate of the southwest entrance is medium. The performances of the three solutions were discussed in details below.

**(1) Increasing the length of the short lane**

The current length of the short shared lane is 212.24 feet, which can accommodate nine vehicles. Now suppose the short shared lane increases into 313.71 feet, which accommodating 14 vehicles. The new design of the intersection is demonstrated in Figure 3.

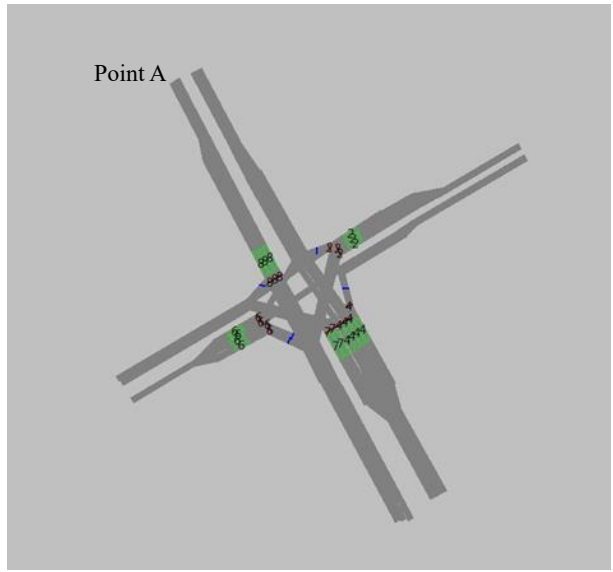

**Figure 3. The intersection with increased length of the short lane.**

It is reasonable to suppose that the vehicles in the left lane will not be affected. For the vehicles in the middle lane and the short lane, the delay times were compared with that of the original intersection. The delay time in VISSIM was calculated as compared to the ideal travel time (free flow condition, no signal control), the mean time delay calculated from all vehicles observed on a single or several link sections (PTV 2011). In this study, the link section of delay time is from the vehicle input point of that link (point A in Figure 3) to the traffic lights location (red bar 8 in Figure 3) at its corresponding lane. The comparison result is shown in Figure 4.

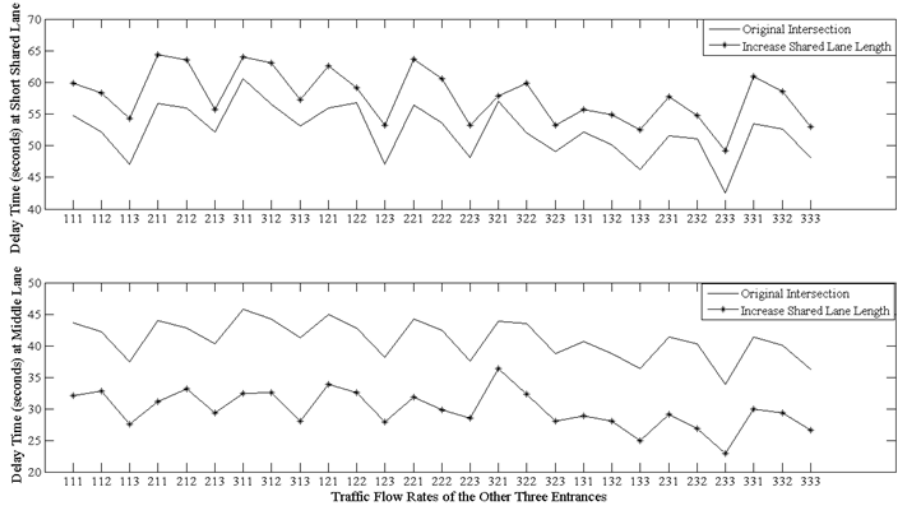

**Figure 4. Delay time comparison of middle and short lanes after increasing the length of the short lane.**

As can be seen, the delay time at middle lane decreases regardless of the 27 situations. However, for the short lane, it shows that the delay time will become longer than the original intersection. The blocking of vehicles at the middle lane may lead to the increased delay time at the short lane. To find out the reason, the maximum queue length at the middle lane is also recorded through VISSIM, and the results are shown as in Figure 5.

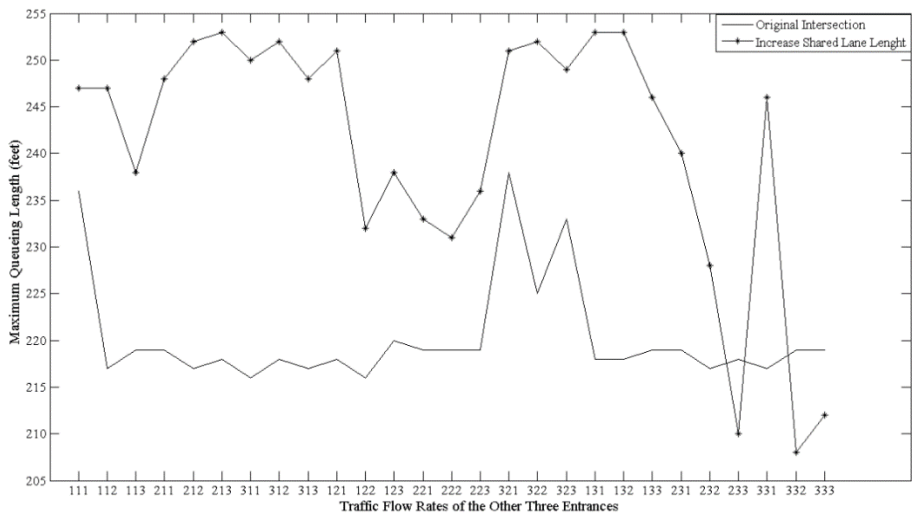

**Figure 5. Maximum queue length changes of middle lane after increasing the short lane length**

Figure 5 shows that as the length of the short shared lane increases, most of the time the maximum queue length is around 250 feet, which is longer than that of the original intersection. However, in VISSIM model, the diverging point of the length increased short lane is 313 feet, which is far away from the queue length detector. It means that the queue at middle lane with a maximum length of around 250 feet will not block the vehicles going to the short shared lane.

Another reason for the increased delay time at short lane is that when the length of the short lane increases, more vehicles will wait in that lane. While in the original intersection model, more vehicles are waiting for the start point of the delay time detector, and in VISSIM this period is not counted as the delay time if the vehicles have not passed the start point (point A in Figure 3). For testing this, the input of vehicles is controlled through VISSIM COM interface (PTV 2011) to make sure the same number of vehicles arrive at the intersection at the same time. Finally, the adjusted delay time at the short lane is shown in Figure 6.

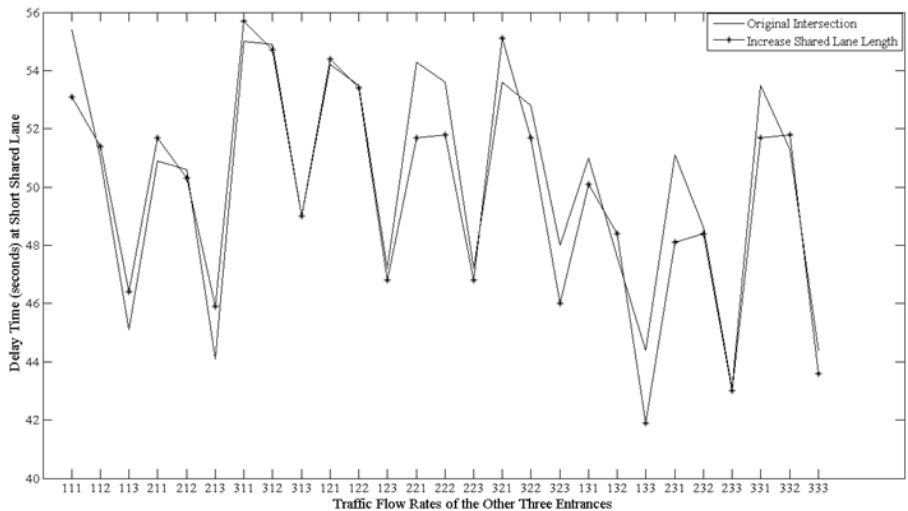

**Figure 6. Delay time comparison of short lane after increasing the short lane length with vehicles input control of VISSIM COM**

**(2) Making the short shared through and right-turn lane as the right-turn-only lane**

Previously it was assumed that 10% of the vehicles at the short lane would turn to the right. Now this solution sets the short shared lane to a right-turn-only lane, so the rest 90% of the vehicles at the short lane need to go through the middle lane. The second column in Table 5 shows the new traffic flow rate for each lane in this scenario I (the traffic flow rate at the middle lane will be the sum of the original traffic flow rate at the middle lane and the transferred flow rate from the short lane). The delay times at the short lane and the middle lane are compared with those of the original intersection in Figure 7.

**Table 5. The New Traffic Flow Rate for Each Lane at Northwest Entrance after Making the Short lane right-turn only (unit: pcu/h)**

| Lanes | scenario I | scenario II |
|---|---|---|
| Left Lane | 401 | 744 |
| Mid Lane | 1095 | 752 |
| Short Lane | 84 | 84 |
| Total | 1,580 | 1,580 |

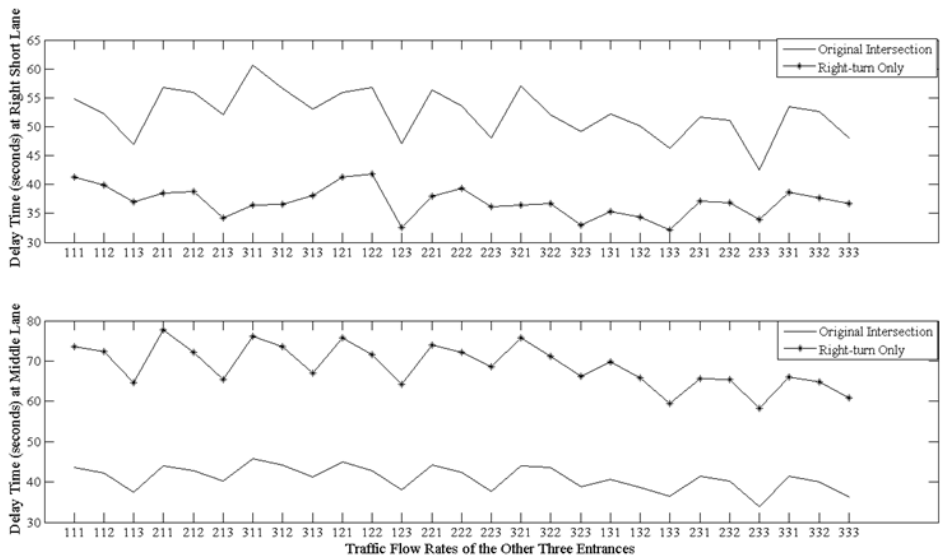

**Figure 7. Delay time comparison of middle and short lanes after making the short lane right-turn only.**

Using this solution, the delay time at the short right lane decreased a lot. However, the delay time at the middle lane increases. It is reasonable because all the vehicles planning to head to the highway are going through the middle lane, and the original vehicles at the middle lane will not change to the left lane, the vehicles heading to the highway are occupying the middle lane. Now assume an extreme scenario II that all the original vehicles at the middle lane will go to the left lane (they may change back to the middle lane after passing the intersection because of the different connected streets downstream), and the vehicles going to the highway go through the middle lane. The new traffic flow rates are shown in the third column in Table 4; the VISSIM model was simulated again. The result is shown in Figure 8.

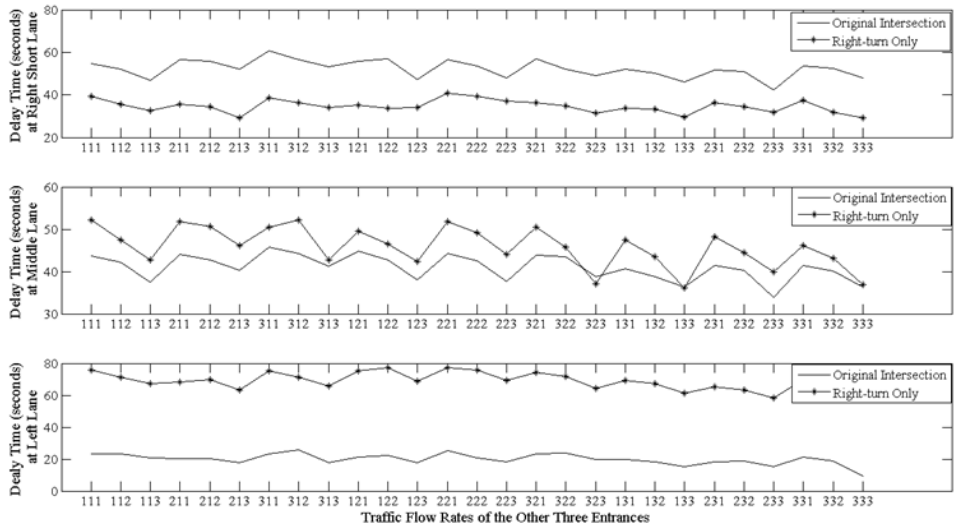

**Figure 8. Delay time comparison of left, middle and short lanes after letting the original vehicles at middle lane go to the left lane.**

If the entire original through traffic changed to the left lane, the delay time at the middle lane is still longer than the original intersection, and the vehicles in the left lane will also experience a much higher delay. Therefore, this solution only improves the quality of service for the right-turn traffic.

**(3) Adding a new diverging lane for the right-turn vehicles**

Solution (3) is to add a diverging link for the right turn vehicles. The new design layout of the intersection in VISSIM is shown in Figure 9.

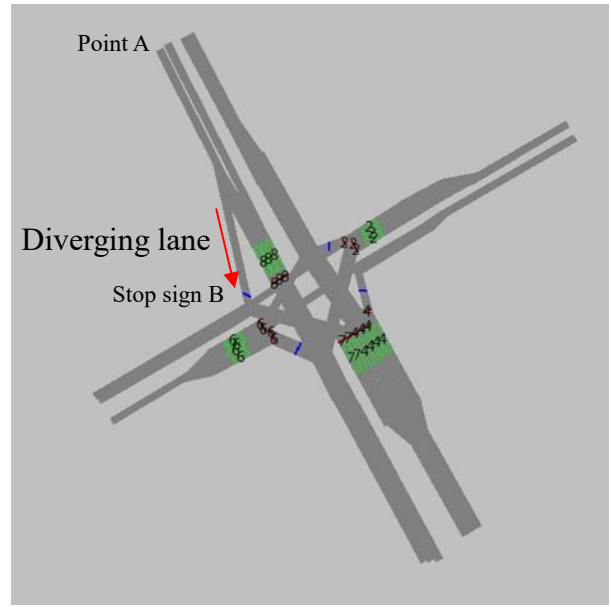

**Figure 9. The new intersection with a diverging lane.**

Then the delay time for the diverging lane (from the vehicle input point A to the stop sign B), short lane, and middle lane are recorded. The comparison with the original intersection is presented in Figure 10.

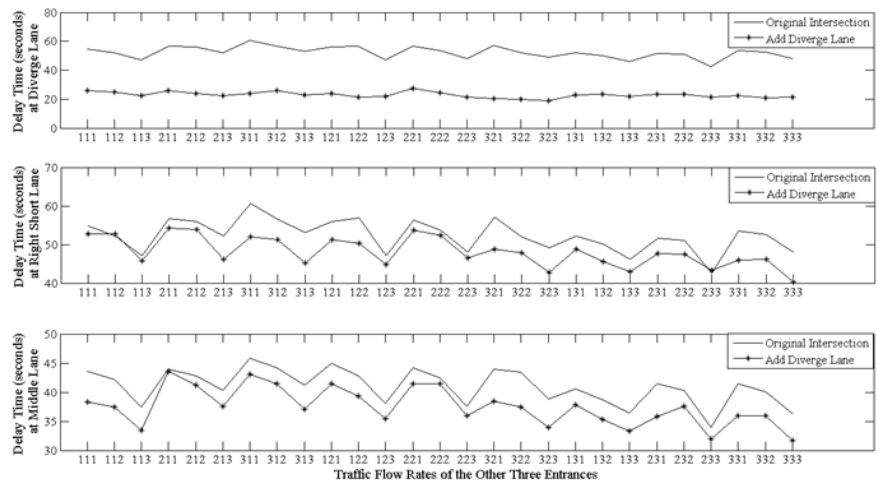

**Figure 10. Delay time comparison of the diverging, short, and middle lanes after adding a new diverging lane for the right-turn vehicles**

The delay time at diverging lane compared with that at the short lane of the original intersection is introduced. It is observed that this solution can prevent the right-turn vehicles being blocked by the vehicles going to the highway. Instead, they can go to West Colvin Boulevard much faster. At the same time, this can save some space at the short lane, which also relieves the pressure to the middle lane for the further step. Therefore, the delay time at the short lane and middle lane are both deceased.

With the help of VISSIM, three solutions improving the current congestion status are observed. More accurate, increasing the length of the short shared through and right-turn lane decreases the delay time at the middle lane, but it has no impacts for the vehicles at the short shared lane. Also, turning the short shared lane as a right-turn-only lane makes the delay time at the short lane shorter, but may cause the delay time at middle and left lanes longer. The third one adding a new diverging lane for the right-turn vehicles performs best which can be beneficial for the right-turn vehicles, as well as the through traffic.

**CONCLUSION AND FUTURE WORK**

This paper proposed three optimal solutions for the long queue problem at one specific intersection caused by the vehicles going to the I-290 highway through the downstream on-ramps. Each solution is tested by building corresponding VISSIM models and compared with the original intersection. A few observations include:

1. An intersection with actuated signal control VISSIM model can be established easily. Besides, VISSIM also provides a few evaluation criteria, like the queue length, delay time;

2. For assessment and comparison of the VISSIM intersection models, the vehicle input should be controlled by COM interface to make sure the number of vehicles arriving at the intersection is equal at the same time;

3. Adding a new diverging link for the right-turn only vehicles can effectively improve the quality of service of this intersection. Meanwhile, the other two solutions, increasing the length of the short shared lane and making the short shared lane right-turn, can only reduce the delay time for the vehicles in one lane.

For the future study, the nearby intersections close to this one will be included in the VISSIM model, and the coordinated signal control will be tested as well. Besides that, this paper focuses on how to improve the layout of the intersection, the configuration of traffic signal could also be considered.

**REFERENCES**

Günther, G., Coeymans, J. E., Muñoz, J. C., and Herrera, J. C. (2012). "Mitigating freeway off-ramp congestion: A surface streets coordinated approach." *Transportation Research Part C: Emerging Technologies*, 20(1), 112–125.

Hunt, P. B., Robertson, D. I., Bretherton, R. D., and Winton, R. I. (1981). "SCOOT - A Traffic Responsive Method of Coordinating Signals." *Publication of: Transport and Road Research Laboratory*.

Li, H., Li, S., Li, H., Qin, L., Li, S., and Zhang, Z. (2016). "Modeling Left-Turn Driving

Behavior at Signalized Intersections with Mixed Traffic Conditions." *Mathematical Problems in Engineering*, Hindawi Publishing Corporation, 2016, 1–11.

Lin, L., Li, Y., & Sadek, A. (2013a). "A k nearest neighbor based local linear wavelet neural network model for on-line short-term traffic volume prediction." *Procedia-Social and Behavioral Sciences*, *96*, 2066-2077.

Lin, L., Wang, Q., & Sadek, A. (2013b). "Short-term forecasting of traffic volume: evaluating models based on multiple data sets and data diagnosis measures." *Transportation Research Record: Journal of the Transportation Research Board*, (2392), 40-47.

Lin, L., Wang, Q., Huang, S., & Sadek, A. W. (2014a). "On-line prediction of border crossing traffic using an enhanced Spinning Network method." *Transportation Research Part C: Emerging Technologies*, *43*, 158-173.

Lin, L., Wang, Q., & Sadek, A. W. (2014b). "Border crossing delay prediction using transient multi-server queueing models." *Transportation Research Part A: Policy and Practice*, *64*, 65-91.

Lin, L., Ni, M., He, Q., Gao, J., & Sadek, A. W. (2015). "Modeling the impacts of inclement weather on freeway traffic speed: exploratory study with social media data." *Transportation Research Record: Journal of the Transportation Research Board, (2482)*, 82-89.

Lin, L., Wang, Q., & Sadek, A. W. (2016). "A combined M5P tree and hazard-based duration model for predicting urban freeway traffic accident durations." *Accident Analysis & Prevention, 91*, 114-126.

Mcelroy, R. (2008). "Traffic Signal Timing Manual."

Mueller, E. A. (1970). "Aspects of the History of Traffic Signals." *IEEE Transactions on Vehicular Technology*.

Niu, H., Li, G., and Lin, L. (2009). "Fuzzy control modeling and simulation for urban traffic lights at single intersection." *Transport Standardization*, 17.

PTV. (2011). *VISSIM 5.40 User Manual*. epubli GmbH, Karlsruhe.

Ring, J. B., and Sadek, A. W. (2012). "Predicting Lane Utilization and Merge Behavior at Signalized Intersections with Auxiliary Lanes in Buffalo, New York." *Journal of Transportation Engineering*, American Society of Civil Engineers, 138(9), 1143–1150.

Tian, Z. Z., and Wu, N. (2006). "Probabilistic Model for Signalized Intersection Capacity with a Short Right-Turn Lane." *Journal of Transportation Engineering*, 132(3), 205–212.

Trabia, M. B., Kaseko, M. S., and Ande, M. (1999). "A two-stage fuzzy logic controller for traffic signals." *Transportation Research Part C: Emerging Technologies*, Elsevier Science Ltd, 7(6), 353–367.

Wu, N. (2007). "Total Approach Capacity at Signalized Intersections with Shared and Short Lanes: Generalized Model Based on a Simulation Study." *Transportation Research Record: Journal of the Transportation Research Board*, 2027(2027), 19–26.